\documentclass[12pt]{iopart}
\usepackage{iopams}
\usepackage{graphicx}

\newcommand{\lr}[1]{ \left( #1 \right) }
\newcommand{\lrs}[1]{ \left[ #1 \right] }
\newcommand{\lrc}[1]{ \left\{ #1 \right\} }

\newcommand{\ket}[1]{ \, | #1 \rangle }
\newcommand{\bra}[1]{ \langle #1 | \, }
\newcommand{\braket}[2]{\langle #1 | #2 \rangle}

\newcommand{\const}{ {\rm const}}

\newcommand{\expa}[1]{ \exp{\left( #1 \right)} }

\newcommand{\logo}{\\ \vskip -18mm
\leftline{\includegraphics[scale=0.3,clip=false]{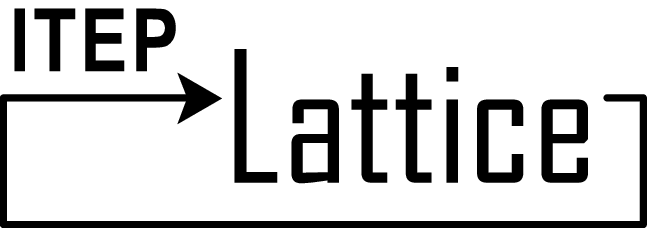}} \vskip 10mm}

\begin{document}
\sloppy

\begin{flushright}
ITEP-LAT/2008-21
\end{flushright}

\title[Entanglement in gauge theories]{Entanglement entropy in lattice gauge theories \logo}

\author{P. V. Buividovich$^{1, 2}$\footnote{Speaker at Confinement8 conference, Mainz, Germany, 1 - 6 September 2008} and M. I. Polikarpov$^{1}$\footnote{Speaker at Liouville Gravity And Statistical Models International conference in memory of Alexei Zamolodchikov, Moscow 21 - 23 June 2008}}

\address{$^{1}$ Institute for Theoretical and Experimental Physics, 117218 Russia, Moscow, B. Cheremushkinskaya str. 25}
\address{$^{2}$ Joint Institute for Power and Nuclear Research "Sosny", 220109 Belarus, Minsk, Acad. Krasin str. 99}
\eads{\mailto{buividovich@tut.by}, \mailto{polykarp@itep.ru}}

\begin{abstract}
 We report on the recent progress in theoretical and numerical studies of entanglement entropy in lattice gauge theories. It is shown that the concept of quantum entanglement between gauge fields in two complementary regions of space can only be introduced if the Hilbert space of physical states is extended in a certain way. In the extended Hilbert space, the entanglement entropy can be partially interpreted as the classical Shannon entropy of the flux of the gauge fields through the boundary between the two regions. Such an extension leads to a reduction procedure which can be easily implemented in lattice simulations by constructing lattices with special topology. This enables us to measure the entanglement entropy in lattice Monte-Carlo simulations. On the simplest example of $Z_{2}$ lattice gauge theory in $\lr{2 + 1}$ dimensions we demonstrate the relation between entanglement entropy and the classical entropy of the field flux. For $SU\lr{2}$ lattice gauge theory in four dimensions, we find a signature of non-analytic dependence of the entanglement entropy on the size of the region. We also comment on the holographic interpretation of the entanglement entropy.
\end{abstract}
\submitto{\JPA}
\maketitle

 The methods of quantum information theory has led recently to many important advances in our understanding of quantum field theories in continuous space-times and on the lattices \cite{Calabrese:06:1}. In particular, quantum entanglement of states of systems with many degrees of freedom turned out to be a very useful model-independent characteristic of the structure of the ground state of quantum fields. A commonly used measure of quantum entanglement of the ground state of quantum fields in $\lr{D-1} + 1$-dimensional space-time is the entropy of entanglement $S\lrs{A}$ between some $\lr{D-1}$-dimensional region $A$ and its $\lr{D-1}$-dimensional complement $B$, which characterizes the amount of information shared between $A$ and $B$ \cite{Calabrese:06:1}. Entanglement entropy is defined as the usual von Neumann entropy for the reduced density matrix $\hat{\rho}_{A}$ associated with the region $A$:
\begin{eqnarray}
\label{ent_def}
S\lrs{A} = - \tr_{A} \lr{ \hat{\rho}_{A} \ln \hat{\rho}_{A} }
\end{eqnarray}
The reduced density matrix is obtained from the density matrix of the ground state of the theory, $\hat{\rho}_{AB} = \ket{0} \bra{0}$, by tracing over all degrees of freedom which are localized outside of $A$, i.e. within $B$ \cite{Calabrese:06:1}:
\begin{eqnarray}
\label{dm_def}
\hat{\rho}_{A} = \tr_{B} \hat{\rho}_{AB} = \tr_{B} \ket{0}\bra{0}
\end{eqnarray}
This density matrix describes the state of quantum fields as seen by an observer who can only perform measurements within $A$.

 The entropy of entanglement of confining gauge theories has recently become a subject of extensive studies in the framework of AdS/CFT correspondence, where a simple geometric expression for the entanglement entropy was conjectured \cite{Ryu:06:1, Takayanagi:06:1, Klebanov:07:1}. One of the most interesting predictions of \cite{Ryu:06:1, Takayanagi:06:1, Klebanov:07:1} is that for confining gauge theories the entanglement entropy (\ref{ent_def}) should be non-analytic in the size of the region $A$. In the limit $N_{c} \rightarrow \infty$ the derivative of the entropy over the size of $A$ changes from being proportional to $N_{c}^{2}$ to a  quantity of order $N_{c}^{0}$. This property can be intuitively understood without any reference to AdS/CFT. Indeed, at small distances an observer in $A$ can only see quarks and gluons, whose number and hence the entropy scales as $N_{c}^{2}$, where $N_{c}$ is the number of colours. At large distances the effective degrees of freedom are quarks and hadrons, whose number and the entropy are of order $N_{c}^{0}$. Thus at some characteristic size of $A$, which should be determined by a typical hadronic scale, the entropy should change from $N_{c}^{2}$ to $N_{c}^{0}$. The nontrivial prediction from AdS/CFT is that this change is stepwise, and in some sense colourless and colourfull degrees of freedom never coexist at one energy scale. A similar non-analuticity has also been predicted using the approximate Migdal-Kadanoff decimations for $SU\lr{2}$ lattice gauge theory \cite{Velytsky:08:1} and observed in lattice simulations \cite{Buividovich:08:2}. Thus the behavior of entanglement entropy of confining gauge theories can be an interesting new test of Maldacena duality between string theories on $\lr{D+1}$-dimensional curved spaces and $D$-dimensional gauge theories which live on their boundary. On the other hand, an attempt to define rigorously what is a quantum entanglement between gauge fields in the regions $A$ and $B$ also leads to the result which closely resembles the holographic principle as formulated by t'Hooft \cite{tHooft:93:1}. Namely, it turns out that due to a specific structure of the Hilbert space of gauge theories, the entanglement entropy includes the classical Shannon entropy of the probability distribution of the flux of the curvature tensor of the gauge field through the boundary $\partial A$ of the region $A$ \cite{Buividovich:08:3}. This entropy can be interpreted as the entropy of the endpoints of electric strings on $\partial A$. Thus the boundary between the regions $A$ and $B$ becomes a sort of $D$-brane for electric strings.

 In order to define the partial trace $\tr_{B}$ in (\ref{dm_def}), one should decompose the Hilbert space of the quantum field theory into a direct product $\mathcal{H}_{A} \otimes \mathcal{H}_{B}$ of Hilbert spaces $\mathcal{H}_{A}$ and $\mathcal{H}_{B}$ of states of quantum fields inside the region $A$ and $B$. One of our main assertions is that such a decomposition is impossible for a Hilbert space of physical states, i.e. a Hilbert space of states which satisfy the Gauss law. The reason is that in contrast to scalar field theories, in pure gauge theories elementary excitations are not associated with points in space, but rather with closed loops, which are the lines of electric flux or electric strings \cite{PolyakovGaugeStrings}. Obviously, closed loops cannot be classified as belonging either to $A$ or $B$, and there are the states which cannot be decomposed into direct products of states localized completely within $A$ or $B$. It should be stressed that this conclusion follows from the general structure of the Hilbert spaces of gauge theories rather than from the dynamical properties of a particular theory, such as confinement or deconfinement.

 For the sake of brevity let us consider the simplest case of $Z_{2}$ lattice gauge theory in $\lr{2 + 1}$ space-time dimensions. All the considerations below can be generalized to other gauge theories with minor modifications. The Hilbert space is a space of all functions $\psi\lrs{ z_{l} }$ of $Z_{2}$-valued link variables $z_{l}$ on a two-dimensional lattice. A convenient basis in this space can be parameterized by a set of integer variables $m_{l} = 0, 1$ defined on the links of the lattice: $\psi\lrs{z_{l}; m_{l}} \sim \prod \limits_{l} z_{l}^{m_{l}}$. This basis is orthonormal with respect to the scalar product $\braket{\psi_{1}}{\psi_{2}} = \sum \limits_{\lrc{z_{l}}} \bar{\psi}_{1}\lrs{z_{l}} \psi_{2}\lrs{z_{l}}$. The Hilbert space of physical states $\mathcal{H}_{0}$ is obtained by imposing the Gauss law as a first-class constraint, so that the wave functions of physical states  are invariant under gauge transformations: $\psi\lrs{ z_{l}^{\Omega} } = \psi\lrs{z_{l}}$. The basis states in $\mathcal{H}_{0}$ can be labelled by such sets $\lrc{m_{l}}$, for which the links with $m_{l} = 1$ form closed non-self-intersecting loops, and can be obtained by acting with all possible Wilson loop operators $W\lrs{C} = \prod \limits_{l \in C} z_{l}$ on the trivial strong-coupling ground state with $\psi\lrs{z_{l}} = \const$. The operator $W\lrs{C}$ thus creates an electric string on the loop $C$.

 Let us denote the Hilbert spaces of all functions on links within $A$ or $B$ as $\tilde{\mathcal{H}}_{A}$ and $\tilde{\mathcal{H}}_{B}$. We would like to decompose the Hilbert space of physical states $\mathcal{H}_{0}$ as $\mathcal{H}_{0} = \mathcal{H}_{A} \otimes \mathcal{H}_{B}$, so that $\mathcal{H}_{A}$ and $\mathcal{H}_{B}$ are the subspaces of $\tilde{\mathcal{H}}_{A}$ and $\tilde{\mathcal{H}}_{B}$.  Consider some state with the wave function of the form $\psi\lrs{z_{l}} \sim \prod \limits_{C} z_{l}$, where the $C$ is some closed loop which belongs both to $A$ and $B$. This state is a direct product of two states in $\tilde{\mathcal{H}}_{A}$ and $\tilde{\mathcal{H}}_{B}$ with wave functions $\psi_{A}\lrs{z_{l}} \sim \prod \limits_{C_{A}} z_{l}$ and $\psi_{B}\lrs{z_{l}} \sim \prod \limits_{C_{B}} z_{l}$, where $C_{A}$ and $C_{B}$ are the parts of the loop $C$ which belong to $A$ and $B$. Since the basis states corresponding to different loops are orthogonal, this decomposition is unique, and the states $\psi_{A}$ and $\psi_{B}$ should necessarily belong to $\mathcal{H}_{A}$ and $\mathcal{H}_{B}$. The spaces $\mathcal{H}_{0}$ and hence $\mathcal{H}_{A}$ and $\mathcal{H}_{B}$ should also contain the trivial strong-coupling ground state with $\psi_{0}\lrs{z_{l}} = \const = \psi_{A \, 0} \psi_{B \, 0}$, $\psi_{A, B \, 0}\lrs{z_{l}} = \const$. Therefore if $\mathcal{H}_{0}$ is indeed a direct product of $\mathcal{H}_{A}$ and $\mathcal{H}_{B}$, it should also contain a direct product of, say, $\psi_{A \, 0}$ with $\psi_{B}$. The wave function for such a state is proportional to $\prod \limits_{C_{B}} z_{l}$, but $C_{B}$ is in general not closed, therefore the Gauss law is violated at the boundary between $A$ and $B$. We thus arrive at the contradiction, which completes the proof.

 Let us now try to extend the Hilbert space of physical states in some minimal way, so that a direct product structure can be introduced. The Hilbert space $\mathcal{H}_{0}$ cannot be decomposed into a direct structure, because it contains the closed strings which belong to both regions $A$ and $B$. We can get rid of such strings, if we allow the strings to open on the boundary between $A$ and $B$. In this case any closed string which crosses $\partial A$ can be represented as a direct product of two open strings, each of which lies completely either within $A$ or $B$. We thus arrive at the minimal extension $\tilde{\mathcal{H}}_{0}$ of $\mathcal{H}_{0}$, which is the Hilbert space of states which violate the Gauss law only at $\partial A$. Minimal extension here means that the Gauss law does not hold in a minimal number of points, and $\tilde{\mathcal{H}}_{0}$ is a subspace of $\mathcal{H}$ of minimal dimensionality which contains $\mathcal{H}_{0}$ and which can be represented as a direct product. To see that this is indeed so, suppose that the Gauss law still holds in some points on $\partial A$. In this case one can repeat all the arguments above for the loop $C$ which goes through these points and arrive at the same contradiction.

 We thus conclude that in order to define the spaces of states of gauge fields which are localized either in $A$ or $B$ one should extend the Hilbert space of physical states and include also the states of electric strings which can open on the boundary $\partial A$ between $A$ and $B$. This boundary can be therefore considered as a sort of $D$-brane for electric strings, with the positions of their endpoints being the additional degrees of freedom which emerge as a result of such extension. It is reasonable to conjecture that these extra degrees of freedom should contribute somehow to the entropy of entanglement between $A$ and $B$.
 In the extended Hilbert space one can naturally define the partial trace over the fields in $B$ as a sum over all links which belong to $B$:
\begin{eqnarray}
\label{reduction}
\bra{z_{l}'} \, \tr_{B} \hat{\rho} \, \ket{z_{l}''} = \sum \limits_{\lrc{z_{l}}, l \in B} \rho\lrs{ \lrc{z_{l}, z_{l}'}, \lrc{z_{l}, z_{l}''}}
\end{eqnarray}
The same expression is valid for lattice gauge theory with any gauge group, with the summation over $Z_{2}$-valued variables $z_{l}$ being replaced by integration over the link variables which belong to the corresponding gauge group.

 Having the reduction procedure (\ref{reduction}) at hand, one can use the standard replica trick, which is commonly used to calculate the entanglement entropy of scalar quantum field theories \cite{Calabrese:06:1}. Namely, we express the entanglement entropy in terms of the free energies $F\lrs{A, s, T} = - \ln \mathcal{Z}\lrs{A, s, T}$ of the theory on a space with topology $\mathbb{C}^{\lr{s}} \otimes \mathbb{T}^{D-2}$, where $\mathbb{C}^{\lr{s}}$ is the $s$-sheeted Riemann surface and $\mathbb{T}^{D-2}$ is the $D-2$ dimensional torus:
\begin{eqnarray}
\label{ent_vs_fe}
S\lrs{A} =
\lim \limits_{T \rightarrow 0}
\lr{
\lim \limits_{s \rightarrow 1} \, \frac{\partial}{\partial s} \, F\lrs{A, s, T}
- F\lr{T}
}
\end{eqnarray}
The branching points of $\mathbb{C}^{\lr{s}}$ should cover the boundary $\partial A$ of $A$.
\begin{center}
\begin{figure}
  \includegraphics[width=6cm]{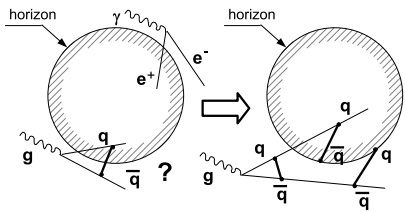}\\
  \caption{Pair production for a confining gauge theory in the vicinity of a black hole.}
  \label{fig:pair_production_BH}
\end{figure}
\end{center}

 In order to show that the endpoints of electric strings on the boundary of $A$ indeed become physically relevant degrees of freedom, let us first consider a confining gauge theory - say, QCD - in the vicinity of a black hole. Since the region inside the horizon is not accessible to an external observer, in order to find the density matrix of the fields outside of the black hole one should trace over all states of fields inside the horizon, which is one possible way to relate the Bekenstein-Hawking entropy with the entanglement entropy. One possible explanation for the radiation from black holes, which is usually ascribed to V. Gribov, is the pair production, in which a photon emits a virtual electron-positron pair which becomes on-shell in the external gravitational field. One of these particles disappears behind the horizon and another is observed as radiated by a black hole. Consider now a similar process, in which a gluon emits a pair of a quark and an antiquark. Say, the quark disappears behind the horizon and an antiquark goes to an external observer. If the Hilbert space of gauge theory in the vicinity of a black hole were constructed in a usual way, we would have to conclude that even after the disappearance of the quark the particles are still connected with a confining string. This, however, would violate the ``no-hair'' theorem. There are only two ways to save the ``no-hair'' theorem: either to assume that a black hole is a source of free quarks, or to assume that a confining string can be torn on the horizon (see figure \ref{fig:pair_production_BH}). The former assumption is obviously prohibited by astrophysical observations, while the latter leads exactly to the extended Hilbert space described above. The horizon of a black hole thus should be a sort of $D$-brane for electric strings of confining gauge theories.

 Consider now a simple trial ground state wave function of $\lr{2 + 1}$-dimensional lattice gauge theory. This trial ground state is a superposition of all possible configurations of closed electric strings with the weight $\expa{ - \frac{\alpha}{2} \, \sum \limits_{l} m_{l}}$, where $\sum \limits_{l} m_{l}$ is the total length of strings:
\begin{eqnarray}
\label{trial_wf}
\Psi_{0}\lrs{z_{l}} = C \sum \limits_{\lrc{ \delta m_{l} = 0}}
\expa{ - \frac{\alpha}{2} \, \sum \limits_{l} m_{l}} \psi\lrs{z_{l}; m_{l}}
\end{eqnarray}
where $\delta m_{l} = 0$ means that we sum only over closed electric strings and $C$ is the normalization constant. The wave function (\ref{trial_wf}) would be a product of functions of individual link variables if the constraint $\delta m_{l} = 0$ was omitted. Thus in some sense the trial ground state (\ref{trial_wf}) is only ``minimally'' entangled because the electric strings are closed, and there is no entanglement between closed string states in the bulk of the regions $A$ and $B$. Hopefully, with such a wavefunction we can calculate the entanglement solely due to the strings which cross $\partial A$.

 After the application of the replica trick and some algebraic manipulations we arrive a the following simple and universal answer for the entanglement entropy of the trial ground state (\ref{trial_wf}):
\begin{eqnarray}
\label{trial_wf_ent_ent}
S\lrs{A} = - \sum \limits_{\lrc{x_{1}, \ldots, x_{m}}} p\lrs{\lrc{x_{1}, \ldots, x_{m}}} \ln p\lrs{\lrc{x_{1}, \ldots, x_{m}}}
\end{eqnarray}
where $p\lrs{\lrc{x_{1}, \ldots, x_{m}}}$ is the probability that the electric string crosses $\partial A$ in the $m$ points $x_{1}, \ldots, x_{m}$. Thus for the trial ground state (\ref{trial_wf}) the entanglement entropy is the classical Shannon entropy of the string endpoints on $\partial A$. This result can be interpreted as follows: an observer within $A$ can not learn whether the electric string are continuous over $\partial A$ or not, and should assume the latter. Although in the ground state all electric strings are, of course, continuous, the observer who can only perform measurements within $A$ can never know this, and for him the positions of string endpoints outside of $A$ are an additional source of uncertainty.

 In order to check the applicability of the expression (\ref{trial_wf_ent_ent}) to the true ground state of $\lr{2 + 1}$-dimensional lattice gauge theory, we have measured both the entanglement entropy and the entropy of string endpoints (\ref{trial_wf_ent_ent}) in Monte-Carlo simulations. In order to measure the entanglement entropy, we have used lattices with topology $\mathbb{C}^{\lr{s}} \otimes \mathbb{T}^{D-2}$ and approximated the derivative over $s$ at $s = 1$ in (\ref{ent_vs_fe}) by a finite difference between $s = 2$ and $s = 1$. The region $A$ was a square of size $l\times l$, and only the differences of entropies $S\lrs{A'} - S\lrs{A} = S\lr{l+1} - S\lr{l}$ were measured.

 The entropy of the endpoints of electric strings was calculated as the Shannon entropy of the classical probability distribution $p\lrs{\lrc{x_{1}, \ldots, x_{m}}}$ of the intersection points $x_{1}, \ldots, x_{m}$ between electric strings and the boundary of $A$ under the simplifying assumption that at fixed $m$ all endpoints are uniformly distributed over $\partial A$. The configurations of electric strings were extracted from lattice simulations of the theory which is related to $Z_{2}$ lattice gauge theory by a Kramers-Wannier duality - namely, the Ising model in $\lr{2 + 1}$ dimensions.

\begin{figure}
  \includegraphics[width=5cm, angle=-90]{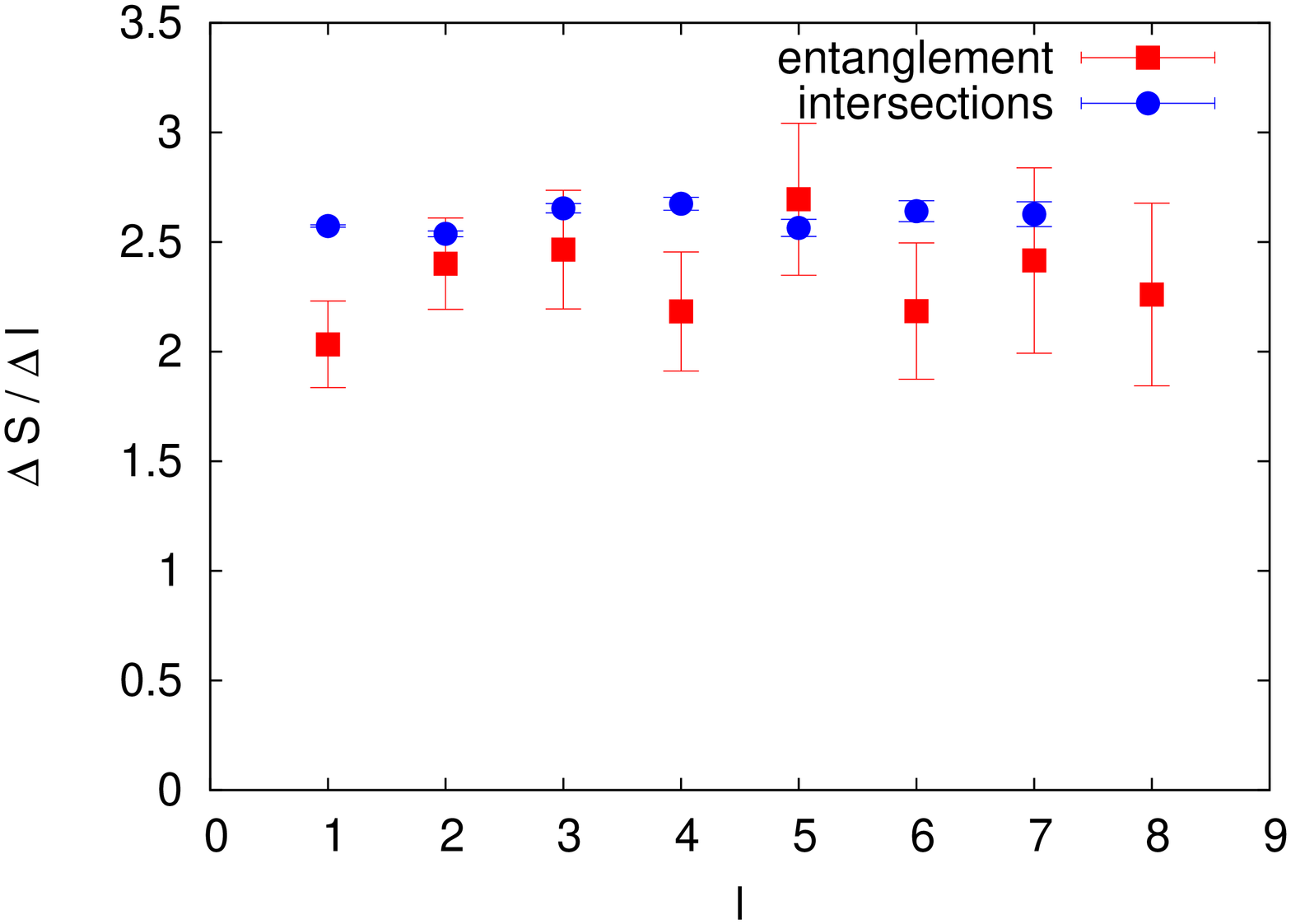} \includegraphics[width=5cm, angle=-90]{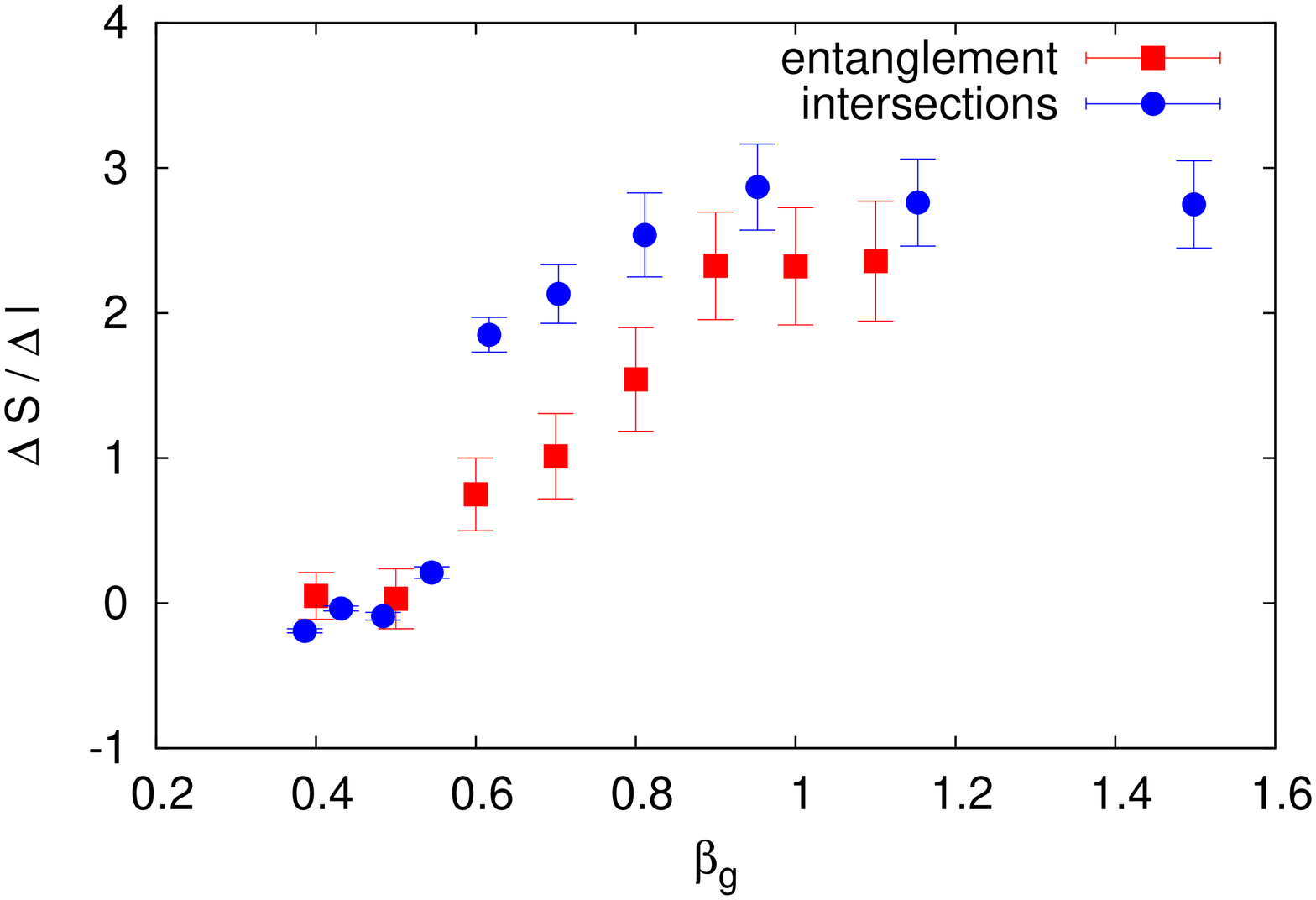}\\
  \caption{Lattice derivatives of the entanglement entropy and the entropy of intersection points over the size of the region for $16^{3}$ lattice. On the left: as a function of $l$ at $\beta_{g} = 0.788$, on the right: as a function of $\beta_{g}$ at $l = 8$.}
  \label{fig:ment_vs_l}
\end{figure}

 Lattice derivatives $S\lr{l+1} - S\lr{l}$ of the entanglement entropy and the entropy of the distribution of intersection points are plotted on figure \ref{fig:ment_vs_l}. For the left plot on figure \ref{fig:ment_vs_l} the coupling constants $\beta_{g} = 0.788$ of $Z_{2}$ lattice gauge theory is close to the critical point at which percolating electric strings emerge. At this value of coupling constant the differences $S\lr{l+1} - S\lr{l}$ of both entropies indeed agree within the error range and practically do not depend on $l$, which means that the entropy is proportional to the area of the boundary with a good precision. We thus recover the familiar ``area law'' for the entanglement entropy. The data plotted on the right plot on figure \ref{fig:ment_vs_l} was obtained for the fixed size of the region $A$ $l = 8$ and for different values of the coupling constants $\beta_{g}$ and $\beta_{s}$. Again, only the differences $S\lr{l+1} - S\lr{l}$ were compared. The dependence on $\beta_{g}$ is again similar for both entropies. We conclude that the entanglement entropy is indeed saturated by the uncertainty in the positions of the endpoints of open electric strings in the extended Hilbert space at the outer side of $\partial A$. In view of the actively discussed ``holographic descriptions'' of field theories, which relate the properties of string theories in the bulk of AdS space or modifications thereof with the properties of the field theories on its boundary, it seems rather tempting to associate the entanglement entropy of the region $A$ with the classical entropy of some statistical theory on its boundary. The endpoints of electric strings could then be interpreted as elementary degrees of freedom in such a theory.

 Finally, let us present the results of the measurements of the entanglement entropy in four-dimensional $SU\lr{2}$ lattice gauge theory. As explained in the introductory paragraph, one can expect that for confining non-Abelian lattice gauge theories the entropy should change rapidly at some scale related to a typical hadronic scale. In order to measure the entanglement entropy in this case, we have used the same approximations as for the $\lr{2 + 1}$ $Z_{2}$ lattice gauge theories considered above.

 \begin{figure}[ht]
  \includegraphics[width=5cm, angle=-90]{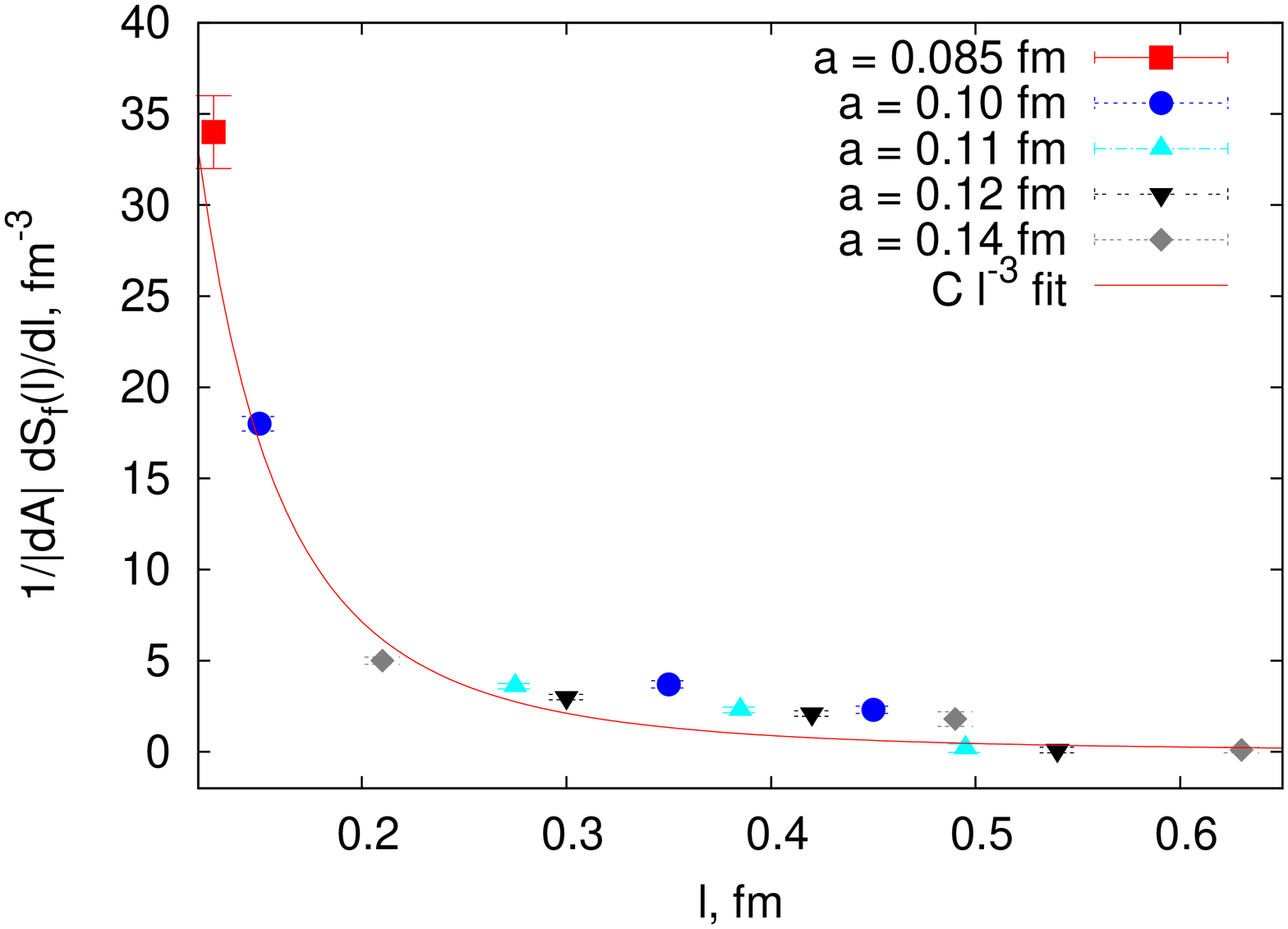}\includegraphics[width=5cm, angle=-90]{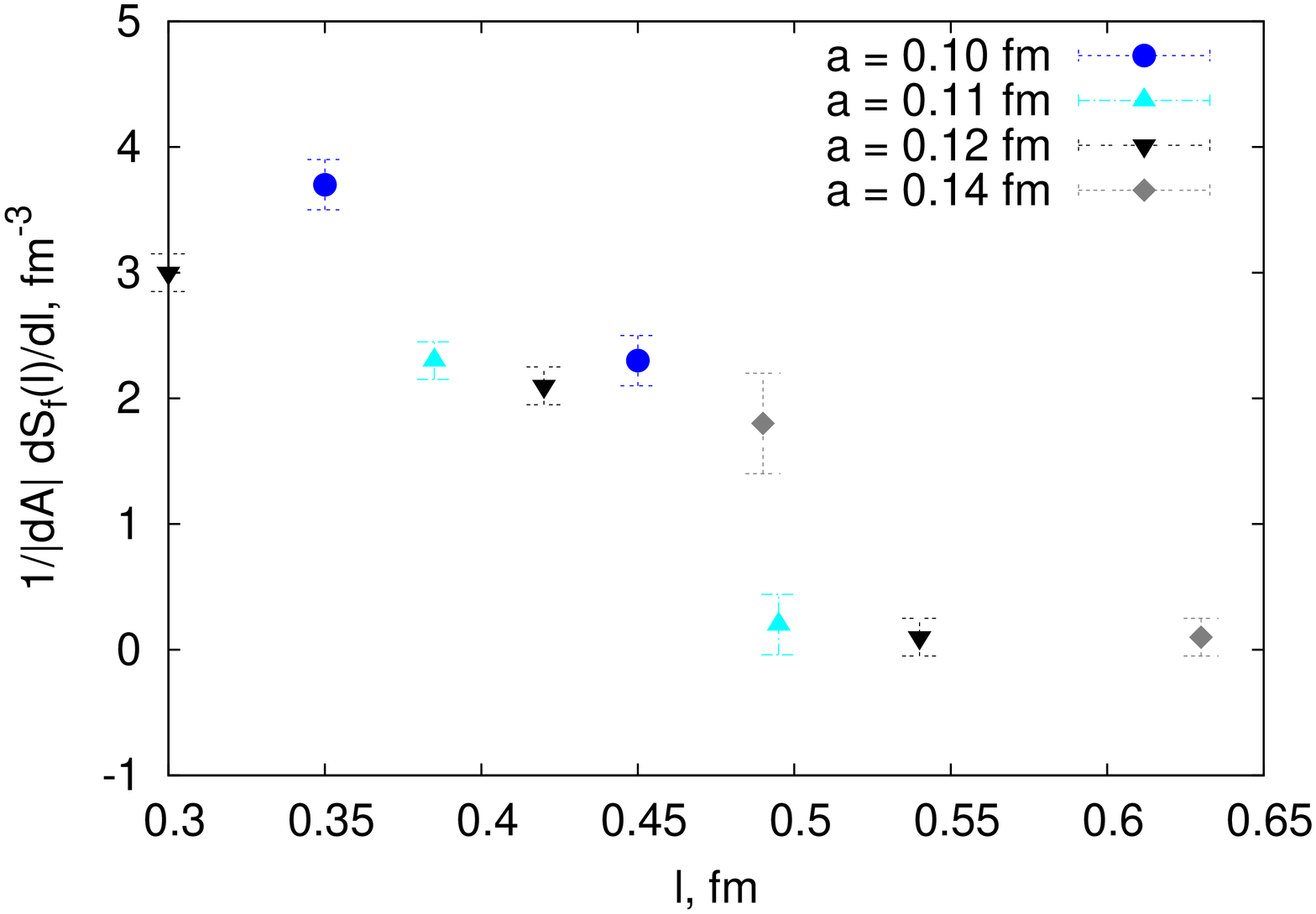}\\
  \caption{The dependence of the derivative of the entanglement entropy $\frac{1}{|\partial A|} \, \frac{S_{f}\lr{l}}{\partial l}$ on $l$. Solid line is the fit of the data by the function $C \, l^{-3}$.}
  \label{fig:Sphys_vs_lphys}
\end{figure}

 The derivative $\frac{1}{|\partial A|} \, \frac{\partial}{\partial l} \, S\lr{l}$ estimated from these measurements is plotted on figure \ref{fig:Sphys_vs_lphys}. It can be seen that this derivative grows rapidly at small distances. For comparison with the asymptotic behavior $\frac{\partial}{\partial l} \, S\lr{l} \sim  l^{-3}$ at $l \rightarrow 0$, we have fitted these results by the function $C l^{-3}$ (solid line on figure \ref{fig:Sphys_vs_lphys}). For the data points with the smallest $l$ the finite differences $a^{-1} \lr{F\lr{l+1, 2, T} - F\lrs{l, 2, T}}$ were found at fixed $l/a$ at different $a$, so that the finite differences $\lr{l + a/2}^{-2} - \lr{l - a/2}^{-2}$ still behave as $l^{-3} \sim a^{-3}$. At larger $l$ $\frac{\partial}{\partial l} \, S_{f}\lr{l}$ seem to approach a kind of plateau for the values of $l$ between $0.3 \, fm$ and $0.5 \, fm$. Here the values of $\frac{\partial}{\partial l} \, S_{f}\lr{l}$ obtained for different values of lattice spacing differ rather significantly, which indicates that for our lattice parameters finite-volume and finite-spacing effects may still be rather strong. Nevertheless, at least qualitatively all data points for different values of $a$ display the same behavior. At $l \approx 0.5 \, fm$ the derivative $\frac{1}{|\partial A|} \, \frac{\partial}{\partial l} \, S\lr{l}$ rapidly goes to zero and remains equal to zero for larger $l$. We thus find an indication of a nonanalytic behavior of entanglement entropy, in accordance with the predictions of \cite{Ryu:06:1, Takayanagi:06:1, Klebanov:07:1, Velytsky:08:1}.

 The entanglement entropy is thus a new and a universal parameter which can be used to characterize the confining and deconfining phases of gauge theories. Interestingly, it should distinguish between confinement and deconfinement even in the presence of dynamical charges and can be therefore considered as a long-sought-for ``order parameter'' for confinement/deconfinement phase transition in QCD. On the other hand,  the entanglement entropy receives an essential contribution from the degrees of freedom on the boundary of $A$. We can interpret this fact in terms of ``holography'' - namely, that the dynamics of gauge theories in $D$ dimensions can be encoded on $D-2$-dimensional manifold $\partial A$. This is at least an interesting point for further research.

\ack

 This work was partly supported by grants RFBR 06-02-04010-NNIO-a, RFBR 08-02-00661-a, DFG-RFBR 436 RUS, grant for scientific schools NSh-679.2008.2 and by Federal Program of the Russian Ministry of Industry, Science and Technology No 40.052.1.1.1112 and by Russian Federal Agency for Nuclear Power. The authors are grateful to E. Akhmedov, V. G. Bornyakov, V. Shevchenko and V. Zakharov for interesting discussions. The part of the calculations have been performed on the MVS 50K at Moscow Joint Supercomputer Center.


\end{document}